\documentstyle[aps,prl,multicol,psfig]{revtex}
\tighten
\begin{document}
\title{Striped phases in the two-dimensional Hubbard model with long-range
       Coulomb interaction}
\author{G. Seibold$^{\dagger}$, C. Castellani$^*$, C. Di Castro$^*$, 
        and M. Grilli$^*$}
\address{$^*$ Istituto Nazionale di Fisica della Materia e
Dipartimento di Fisica, Universit\`a di Roma ``La Sapienza'',\\
Piazzale A. Moro 2, 00185 Roma, Italy}
\address{$^{\dagger}$ Institut f\"ur Physik, BTU Cottbus, PBox 101344, 
         03013 Cottbus, Germany}
\date{\today}
\maketitle

\begin{abstract}
We investigate the formation of partially filled domain
walls in the two-dimensional Hubbard model in the presence of 
long-range interaction. Using an unrestricted Gutzwiller variational
approach we show that: i) the strong local interaction favors
charge segregation in stripe domain walls; ii) 
The  long-range interaction favors the
formation of half-filled vertical stripes with
a period doubling due to the charge and a period quadrupling
due to the spins along the wall. 
Our results show that, besides the underlying lattice structure,
also the electronic interactions can contribute to determine
the different domain wall textures in Nd doped copper
oxides and nickel oxides.
\end{abstract}

\vspace*{0.2cm}


\begin{multicols}{2}
The occurence of charged domain walls in the high-T$_{c}$ superconductors
presently attracts a lot of interest also with regard
to possible pairing scenarios \cite{CHINA,CAST,EMERY}.
Incommensurate spin correlations have first been observed in 
La$_{2-x}$Sr$_{x}$CuO$_{4}$ (LSCO)  
 by neutron scattering experiments \cite{CHEONG,MASON,THURST}.
More recently it was found that the incommensurate spin fluctuations
are pinned in Nd doped LSCO and nickel oxide compounds 
\cite{TRAN0,TRAN,TRAN2}
leading to spin- and charge-stripe order in these materials
\cite{notaBIANCONI}.
However, whereas the domain walls in the hole-doped
La$_{2}$NiO$_4$ system  are oriented along the diagonals of
the NiO$_2$ lattice, it turns out that in 
La$_{1.48}$Nd$_{0.4}$Sr$_{0.12}$CuO$_{4}$ the orientation of the
stripes is along the Cu-O bond direction. 
Moreover, the hole
concentration in the domain walls is one hole per Ni site in the
nickelates and one hole every second Cu site in the Nd-doped
cuprates. A comparison of the
low-temperature orthorhombic  and the low-temperature tetragonal
structure suggests that the vertical stripes in 
La$_{1.48}$Nd$_{0.4}$Sr$_{0.12}$CuO$_{4}$ are pinned by the 
tetragonal lattice
potential whereas the orthorombic phase in the nickel oxides favors a
diagonal orientation.  We will show here that the electronic 
interactions may also play an important role in establishing 
different domain wall structures.

The stripe instability for doped antiferromagnets 
was predicted theoretically in \cite{ZAANEN} within 
Hartree-Fock (HF) theory applied
to the extended Hubbard model and confirmed by a number 
of subsequent investigations 
\cite{VARIOUS}.
For small values of the Hubbard on-site repulsion U
(generally smaller than 3$t$ - 4$t$) these calculations result
in a striped phase oriented along the (10)- or (01)-direction
whereas for higher values of U the orientation is along
the diagonals. Within HF theory the stripe solutions become
unstable for $U >8t$ towards the formation of isolated spin-polarons.

However, all stripe calculations performed so far within the HF
approximation of the Hubbard model predict one hole per site
along the domain wall (filled stripe) \cite{WHITE}. 
This contrasts the observation of half-filled stripes
(i.e. with half a hole per site)
in the La$_{1.6-x}$Nd$_{0.4}$Sr$_{x}$CuO$_{4}$ system
which is a yet unresolved problem of mean-field theory.
In Ref.\ \cite{OLES} it was addressed the question
whether the inclusion of an additional nearest neighbor repulsion V 
in the Hubbard-hamiltonian may favor the formation of partially filled
stripes. According to Ref. \onlinecite{OLES}
the half-filled stripe solution
is stabilized by a quadrupling of the charge- or spin-period along the stripe.
However, although the nearest neighbor repulsion slightly
enhances the stability of the half-filled wall this never corresponds
to the HF ground state for realistic parameter values.
Instead the main effect of V is to shift the crossover to 
isolated spin polarons to lower values of U. 

In the present paper we show that a proper treatment of the
strong local repulsion $U$ plays an indirect but crucial role
in stabilizing half-filled vertical domain walls. Specifically we
apply a slave-boson version of the Gutzwiller approach
within an unrestricted variational scheme. Contrary to the pure HF approach,
which heavily underestimates the effective attraction between the charge 
carriers and predicts repulsion for $U>6t$, it was recently shown that
within the slave-boson scheme the attraction persists up to very
large $U$ \cite{GSH}. As a consequence of this more suitable treatment of the
strong coupling limit, $U$ greatly favors the charge segregation
in striped domains as opposed to  spin polarons. 
In the absence of long-range (LR) forces, completely
filled diagonal stripes stay more stable than half-filled 
vertical ones. However, due to their increased stability
with respect to isolated polarons, 
the stripe solutions now allow for a less disruptive
introduction of stronger LR forces, 
which  affect  the completely filled stripes more than the
half-filled ones. Then, for a sizable but still realistic
LR repulsion, the half-filled vertical stripe may become the ground-state
configuration. 

We consider the two-dimensional Hubbard model on a square lattice, with 
hopping restricted to nearest neighbors (indicated by the bracket $<i,j>$)
and an additional LR interaction:
\begin{equation}\label{HM}
H=-t\sum_{<ij>,\sigma}c_{i,\sigma}^{\dagger}c_{j,\sigma} + U\sum_{i}
n_{i,\uparrow}n_{i,\downarrow}+\sum_{i\not= j,\sigma\sigma'}V_{ij}
n_{i,\sigma}n_{j,\sigma'}
\end{equation}
where $c_{i,\sigma}$ destroys an electron 
with spin $\sigma$ at site
i, and $n_{i,\sigma}=c_{i,\sigma}^{\dagger}c_{i,\sigma}$. $U$ is the
on-site Hubbard repulsion and $t$ the transfer parameter. For all calculations
we take $t=1$. For the Coulomb potential we assume an interaction of the 
form $V_{ij}=\frac{V_0}{\sqrt{({\bf R_i}-{\bf R_j})^2 + \alpha^2}}$
where the parameters $V_{0}$ and $\alpha$ are specified through the
on-site repulsion U and the nearest-neighbor interaction $V_{n,n+1}$
via $\alpha=V_{n,n+1}/\sqrt{U^2-V^2_{n,n+1}}$ and $V_0=\alpha U/2$.
Since we consider a finite lattice with periodic boundary
conditions it is also necessary to restrict the LR
part to half of the lattice dimension both in x- and y-direction.

Following Kotliar and Ruckenstein \cite{KOTLIAR} we enlarge the original 
Hilbert space by introducing four subsidiary boson fields $e_{i}$, 
$s_{i,\uparrow}$, $s_{i,\downarrow}$, and $d_{i}$ for each site i. 
These operators stand for the annihilation of
empty, singly occupied states with spin up or down, and doubly occupied 
sites, respectively. Since there are only four possible states per site, 
these boson projection operators must satisfy the completeness
constraints 
$e_{i}^{\dagger}e_{i}+\sum_{\sigma}s_{i,\sigma}^{\dagger}s_{i,\sigma}
+d_{i}^{\dagger}d_{i}=1$ 
and 
$n_{i,\sigma}=s_{i,\sigma}^{\dagger}s_{i,\sigma}+d_{i}^{\dagger}d_{i}$.
In the saddle-point approximation, all bosonic operators are treated
as numbers. Furthermore we will approximate the LR part
by a HF decoupling.
The resulting effective one-particle Hamiltonian 
describes the dynamics
of particles where the hopping amplitude between states
(i,$\sigma$) and (j,$\sigma$) is renormalized by a factor 
$z_{i,\sigma}^\dagger
z_{j,\sigma}$ with
$z_{i,\sigma}=\left( e_{i}^2+s_{i,-\sigma}^2 \right)^{-1/2} (e_{i}s_{i,\sigma}
+s_{i,-\sigma}d_{i})\left( d_{i}^2+s_{i,\sigma}^2 \right)^{-1/2}$
The effective one-particle Hamiltonian
 can be diagonalized by the transformation
$c_{i,\sigma}=\sum_{k}\Phi_{i,\sigma}(k)a_{k}$
where the fermion wave functions  $\Phi_{i,\sigma}(k)$
obey the orthonormality constraint
$
\sum_{i,\sigma}\Phi^{\ast}_{i,\sigma}(k)\Phi_{i,\sigma}(q)=\delta_{kq}.
$

Given a system with $N_{el}$ particles we obtain for the
total energy
\begin{eqnarray}
E_{tot}&=&-t\sum_{<ij>,\sigma}z_{i,\sigma}^{\ast}z_{j,\sigma}
\sum_{k=1}^{N_{el}}
\Phi^{\ast}_{i,\sigma}(k)\Phi_{j,\sigma}(k)+U\sum_{i}d_{i}^{2}\nonumber \\
&+&\sum_{i\not= j,\sigma\sigma'}V_{ij} \sum_{k,k'}\Phi^{\ast}_{i,\sigma}(k)
\Phi_{i,\sigma}(k)\Phi^{\ast}_{j,\sigma'}(k')\Phi_{j,\sigma'}(k')
\label{E1}
\end{eqnarray}
which has to be minimized with respect to the fermionic wavefunctions
and bosonic fields
within the orthonormality and completeness constraints
(for further details of this approach see \cite{GSH}).

{\it Single-stripe calculation. -- }
In the presence of LR interactions a single stripe will always result
unstable with respect to isolated polarons by increasing the length
of the stripe. Indeed the Coulomb energy per hole of a charged
wall of length $L$ increases as $\log L$ and it is not compensated 
by the coupling to a uniform distribution of background charges 
of opposite sign, as it is the case for a regular array of
stripes. Nevertheless the analysis of the single-stripe case
allows to illustrate the stabilization effect of a
large $U$ (within the Gutzwiller approach and with respect to
the HF treatment) and to find out the most stable stripe configuration
within each class of stripes.

It has been intensively discussed in
Ref.\ \cite{OLES} that for the half-filled wall to become
a saddle-point for any effective one-particle description,
it is necessary to quadruple the period along the stripe.
In fact there exist several possibilities of performing this
period quadrupling and in Fig.\ 1 we sketch the charge-
and spin structures of the two types of half-filled vertical
domain walls which have the lowest energy among the different
realizations.
We find that the stripe configurations investigated
in Ref. \cite{OLES} (cf. Figs. 8,9 therein) 
are always higher in energy than the structures shown in Fig.\ 1.  
\begin{figure}
{\hspace{0.3cm}{\psfig{figure=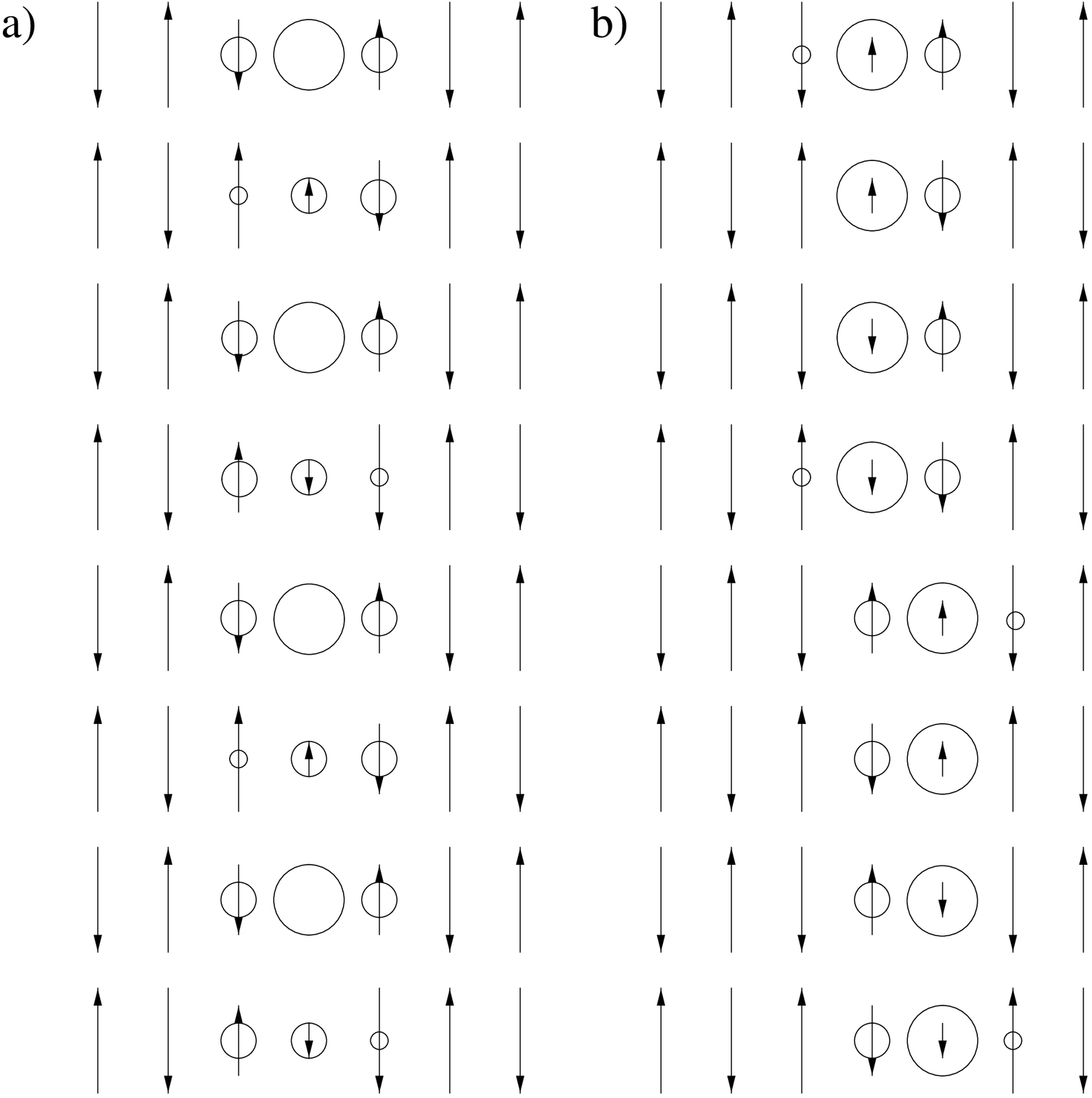,width=6.0cm,angle=-90}}}

{\small FIG. 1. Charge- and spin-density structure for the two half-filled
         vertical domain wall types described in the text. The diameter
         of the circles indicates the hole charge and the length of the spin
         arrows symbolizes the value of $\langle S_z\rangle$.}
\end{figure}
Fig.\ 1(a) is of the CDW type where the main charge modulation is
with $q_{\|}^{c}=\pi$ along the wall whereas
the spin varies with $q_{\|}^{s}=\pi/2$. We find that 
among the half-filled vertical walls this configuration
is lowest in energy in the range
$4t < U < 7.5t$.  However, for strong on-site
repulsion ($U > 7.5t$) the charge and spin realization with the
lowest energy  corresponds to
the staggered structure indicated in Fig.\ 1(b) although the difference
in energy to the CDW-type of Fig.\ 1(a) is rather small 
(the difference in energy per hole for $U=10t$ is 
$\approx 5*10^{-3}t$).

For the single
stripe solutions the size of the supercell is $9\times 8$.
Calculation of the diagonal structures for
periodic boundary conditions without frustrating the 
background spin configuration in principle requires
an 'uneven $\times$ uneven' lattice which
does not allow for the investigation of half-filled
diagonal walls.
For this reason and to avoid the comparison between different lattice 
sizes for different domain wall types we have
choosen supercells with periodic boundaries for vertical stripes
and with shifted periodic boundaries for diagonal stripes.
In the 'shifted boundaries' the
equivalent supercells in the vertical direction
are horizontally shifted by one lattice spacing
(cf. Fig.\ 4 in Ref. \cite{GSH}).  To avoid double counting in the
Coulomb interaction energies, we cut off the LR forces
at ``half-minus-one'' the size of the supercell.

The energy {\it per hole} of each configuration was calculated with respect
to the reference state of a uniform AF lattice with one particle per site:
$E(config)=\left( E(N_h)-E_{AF}(0) \right)/N_h$ \cite{notaenergy}.
In particular the energy of an isolated spin-polaron is given by
$E(polaron)=E(N_h=1)-E_{AF}(0)$.
To compare the stability of stripes on our finite-size system, we
report the energy differences $E(config)-E(polaron)$.

The results are displayed in Fig.\ 2 
as a function of the nearest neighbor
value of the Coulomb repulsion $V_{n,n+1}$
for $U=7t$ in Fig. 2(a) and for $U=10t$ in Fig. 2(b).
In case of half-filled vertical stripes the curves in
Fig.\ 2(a) and 2(b) correspond to the structures in 
Fig.\ 1(a) and 1(b) respectively.
\begin{figure}
\hspace{1.5cm}{{\psfig{figure=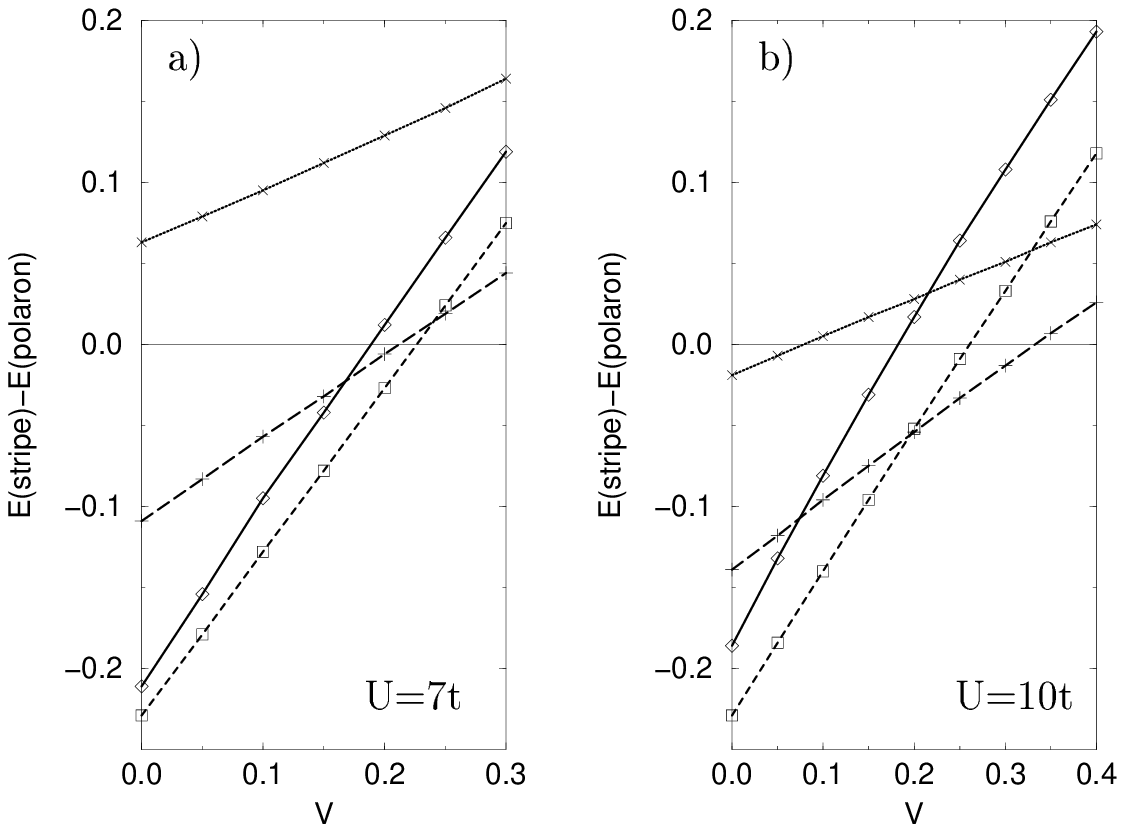,width=6cm}}}

\vspace*{0.7cm}

{\small FIG. 2. Energy difference of various striped phases and the isolated
         polaron lattice $E(stripe)-E(polaron)$ as a function
         of the nearest neighbor contribution $V_{n,n+1}$
	 to the long-range
         potential. Solid line ($\Diamond$): completely filled
         vertical stripe; long-dashed line ($+$): half-filled vertical
         stripe; 
         short-dashed line ($\Box$): completely filled diagonal stripe; 
         dotted line ($\times$): half-filled diagonal stripe. 
         System size: $9\times 8$; a) U=7t; b) U=10t.} 
\end{figure}
Disregarding the eventual above-mentioned instability of 
the single stripes with respect to isolated polarons
by increasing lenght, various features are worth noting.
First of all, completely filled stripes 
increase more rapidly their energy (have a larger slope)
than half-filled ones upon increasing the LR repulsion.
Therefore LR forces favor half-filled stripes,
which eventually become the most favorable wall textures.

However, the most relevant effect to be noticed here 
is the role of a large local repulsion $U$ affecting the
energies of the various textures. In particular
a comparison between Fig. 2(a) and Fig. 2(b) shows that
$U$ strongly reduces the energy of the half-filled
stripes with respect to the filled ones already at $V=0$.
By combining this reduction with the effects of LR forces
on the stripes it follows that increasing $U$
makes the half-filled vertical stripe
the most stable among the wall solutions at smaller
values of $V_{n,n+1}$. 
%
%

{\it Interstripe interaction and spin-polaron lattice. -- }
To assess the actual ground state we now consider the 
interaction between stripes or between spin polarons. 
Indeed one
cannot neglect the interstripe repulsion since,  for example, in 
La$_{1.48}$Nd$_{0.4}$Sr$_{0.12}$CuO$_{4}$ the stripe 
separation is four times the Cu-Cu distance only.
To incorporate the repulsion between stripes, we have calculated
the energy of vertically oriented stripes on a $32 \times 4$ lattice.
For a concentration of $1/8$ this results in an array of 4 completely
filled or 8 half-filled stripes. The 
energy of these arrays with respect to 
the polaronic Wigner lattice are depicted
in Fig.\ 3 (the Wigner lattice now
corresponds to 16 spin polarons with maximum distance). 
\begin{figure}
\hspace{0.1cm}{{\psfig{figure=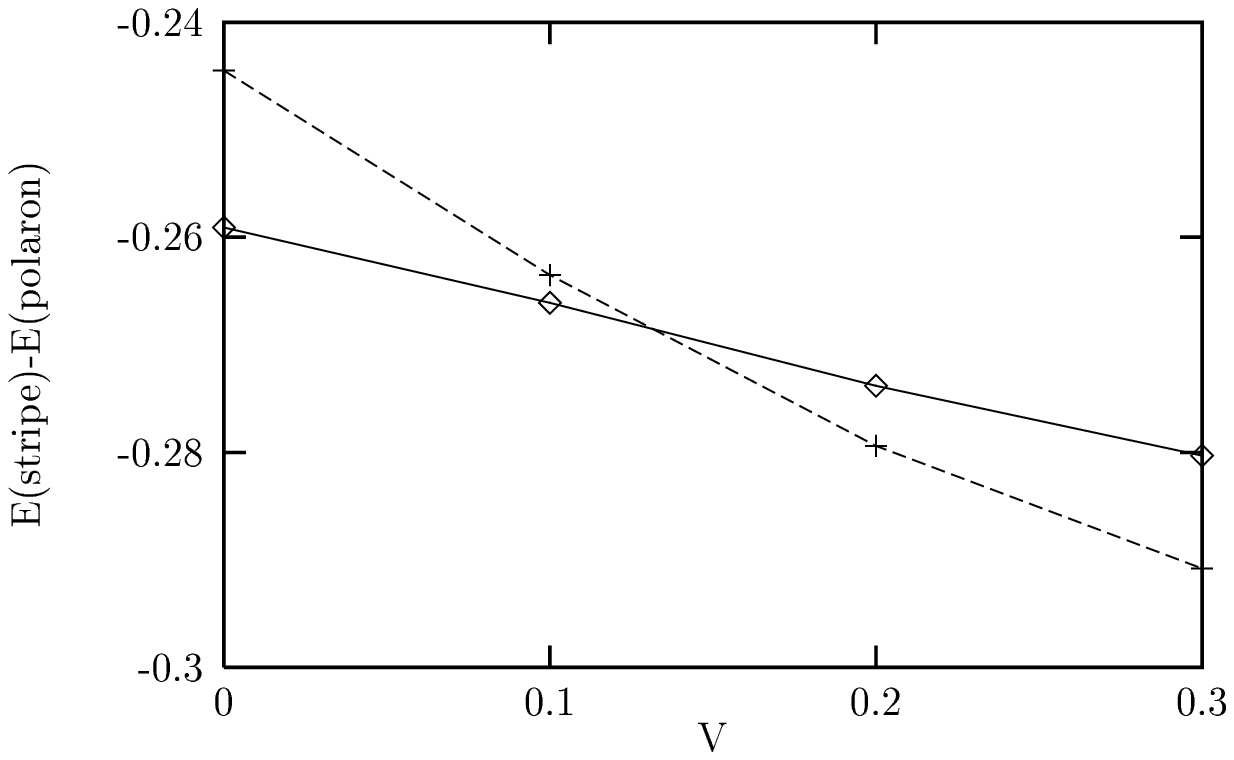,width=7.5cm}}}

{\small FIG. 3. Energy difference between vertically striped phases 
         and the 'interacting' spin-polaron Wigner lattice as a function
         of the nearest neighbor contribution $V_{n,n+1}$
	 to the long-range
         potential (in units of $t$).
         Solid line ($\Diamond$): completely filled
         vertical stripe; long-dashed line ($+$): half-filled vertical
         stripe. System size $32 \times 4$, U=10t.} 
\end{figure}
In this case we obtain a crossover
to half-filled vertical domain walls for $V\approx 0.12t$.
It is interesting to observe that now the stripe
solutions gain in energy upon switching on the LR part
with respect to the polaron lattice. This enhances the parameter
range of stability for the stripes, which
 no longer become unstable towards the decay into isolated polarons. 

To assess the absolute stability of the half-filled
vertical stripes, we should also compare their energy with
the filled diagonal stripes, which, in the single-stripe analysis
result to be more stable than the filled vertical stripes.
However, filled diagonal stripes are strongly 
destabilized by the elongated shape of the $32\times 4$ supercell
so that we do not included their energy in Fig. 3.
As an alternative to the direct calculations on the
elongated cluster, to extract informations about diagonal
stripe configurations, 
 we analyzed a twodimensional regular array of charged
wires with fixed global charge density. We have found that the
electrostatic potential energy is lower for  
wires with higher linear charge density
at a larger distance than for less charged wires more closely spaced.
Therefore diagonal stripes, which at given
planar density are closer by $\sqrt{2}$,
but less densely charged by the same factor, are less favorable than the
vertical stripes as far as the electrostatic Coulombic energy is
concerned. On the other hand, our single-stripe investigation 
already demonstrated that a proper treatment 
of the strong local repulsion $U$
opens the way to a stabilization of half-filled vertical stripes
with respect to the filled diagonal stripes. From the above
 purely electrostatic analysis and
from the results of Fig. 3 we can therefore safely conclude 
that half-filled vertical stripes are the ground-state
configuration for $V \gtrsim 0.12 t$ at large enough $U$.
 
To summarize, we have shown that a LR Coulomb interaction
added to the 2D Hubbard model gives rise to 
half-filled vertically oriented 
domain walls when treated within an unrestricted Gutzwiller approach. 
This feature does not appear in
semiclassical Hartree-Fock approximations where the effective attraction
between spin-polarons is underestimated. 
Depending on the value of the on-site repulsion U,
we expect the domain wall structure in the Nd-doped LaCuO
system to be of the type shown in Fig.\ 1(a)
and Fig.\ 1(b), respectively.
Moreover, the here investigated competition between
completely filled diagonal and half-filled vertical stripes
can explain the different hole orderings in nickelates and in Nd
cuprates even without invoking a relevant role of lattice interactions.
Specifically, our findings suggest that 
nickelates could be characterized by a
smaller $U\lesssim 7t$ and/or a 
smaller $V$ accounting for their filled diagonal
stripes. On the other hand, 
although the half-filled vertical stripes in
the Nd-enriched LaSrCuO systems are to some extent likely fixed along
the (1,0) direction by the underlying lattice structure, we
showed here that, if
large values of $U$ and sizable values of $V$
characterize these systems, then
 electronic correlations would also contribute to give rise to
vertically half-filled stripes.

In general the filling and orientation of the stripes depend
on the specificity of the electronic forces and structures
and therefore inside the various oxide families
different textures of the stripe phase may prevail. 

G.S. acknowledges financial support from the Deutsche Forschungsgemeinschaft
as well as  hospitality and support from the Dipartimento
di Fisica of Universit\`a di Roma ``La Sapienza''where part of this work
was carried out. This work was partially supported by INFM-PRA (1996).

\end{multicols}


\begin{references}
\bibitem{notaenergy} The $E_{AF}$ was calculated for an $8\times 8$
           cluster and rescaled by a factor $(9\times 8)/(8 \times 8)$.
\end{references}
\end{document}